\documentclass{SCGE}
\usepackage{graphicx}
\usepackage{rotating}
\begin{document}

%%%%%%新版式要加上这组
\begin{picture}(0,0){\rm
\put(0,-20){\makebox[160truemm][l]{\bf {\sanhao\raisebox{2pt}{.}}
News and Views  {\sanhao\raisebox{1.5pt}{.}}}}}
%\put(0,-34){\jiuwuhao {\textcolor[rgb]{0.5,0.5,0.5}{\sf Progress of Projects Supported by %NSFC}}}%%(11??×￠êí￡oμ÷\textcolor[rgb]{x,x,x}?Dμ?êy×?x??′ó???ò)
\end{picture}

\def\bm{\boldsymbol}

\def\dl{\displaystyle}
\def\du{\end{document}}
\def\d{{\rm d}}
\def\e{{\rm e}}
\def\i{{\rm i}}

\def\pi{{\uppi}}

% The author doesn't need fill in it.
\Year{2018} %
\Month{October} %
\Vol{xx} %  ?ío?
\No{xx} %  ?úo?
\BeginPage{1} % ?eò3??
\AuthorMark{{\rm Wang}}  %(11??×￠êí￡oò3??é?μ?×÷??)
\AuthorMarkCite{{\rm Wang}. } %(11??×￠êí￡ocitation?Dμ?×÷??)
\DOI{...} % The author doesn't need fill in it.
\ArtNo{xx}

% \title[short text for running head]{full title}{comments for title}
\title[Pulsar giant pulse]{Pulsar giant pulse: coherent instability near light cylinder}

\author[1,2,3]{Weiyang WANG}{}
\author[4,5]{Jiguang Lu}{}
\author[6,7,8,9]{Songbo ZHANG}{}
\author[1,2,10]{Xuelei CHEN}{}
\author[11,12]{Rui LUO}{}
\author[3,11]{Renxin XU}{}

\address[1]{Key Laboratory for Computational Astrophysics, National Astronomical Observatories, Chinese Academy of Sciences, 20A Datun Road, Beijing 100012, China}
\address[2]{University of Chinese Academy of Sciences, Beijing 100049, China}
\address[3]{School of Physics and State Key Laboratory of Nuclear Physics and Technology, Peking University, Beijing 100871, China}
\address[4]{National Astronomical Observatories, Chinese Academy of Sciences, Beijing 100012, China}
\address[5]{Key Laboratory of Radio Astronomy, Chinese Academy of Science, Beijing 100012, China}
\address[6]{Purple Mountain Observatory, Chinese Academy of Sciences, Nanjing 210008, China}
\address[7]{University of Chinese Academy of Sciences, Beijing 100049, China}
\address[8]{CSIRO Astronomy and Space Science, Australia Telescope National Facility, Box 76 Epping NSW 1710, Australia}
\address[9]{International Centre for Radio Astronomy Research, University of Western Australia, Crawley, WA 6009, Australia}
\address[10]{Center for High Energy Physics, Peking University, Beijing 100871, China}
\address[11]{Kavli Institute for Astronomy and Astrophysics, Peking University, Beijing 100871, China}
\address[12]{Department of Astronomy, School of Physics, Peking University, Beijing 1008}

\maketitle \vspace{-3.5mm}{\footnotesize\begin{center}
Received February 13, 2018; accepted March 7, 2018; published online May 20, 2018
\end{center}}\vspace*{-5mm}

%     Abstract is required.
\begin{center}
\rule{16.5cm}{0.4pt}
\parbox{16.5cm}
{\begin{abstract}
Giant pulses (GPs) are extremely bright individual pulses of radio pulsar. In microbursts of Crab pulsar, which is an active GP emitter, zebra-pattern-like spectral structures are observed, which are reminiscent of the ``zebra bands'' that are observed in type IV solar radio flares. However, band spacing linearly increases with the band center frequency of $\sim5-30$\,GHz.
In this study, we propose that the Crab pulsar GP can originate from the coherent instability of plasma near a light cylinder. Further, the growth of coherent instability can be attributed to the resonance observed between the cyclotron-resonant-excited wave and the background plasma oscillation. The particles can be injected into the closed-field line regions owing to magnetic reconnection near a light cylinder.
These particles introduce a large amount of free energy that further causes cyclotron-resonant instability, which grows and amplifies radiative waves at frequencies close to the electron cyclotron harmonics that exhibit zebra-pattern-like spectral band structures. Further, these structures can be modulated by the resonance between the cyclotron-resonant-excited wave and the background plasma oscillation.
In this scenario, the band structures of the Crab pulsar can be well fitted by a coherent instability model, where the plasma density of a light cylinder should be $\sim10^{13-15}\,\rm{cm^{-3}}$, with an estimated gradient of $>5.5\times10^5\,\rm{cm^{-4}}$. This process may be accompanied by high-energy emissions. Similar phenomena are expected to be detected in other types of GP sources that have magnetic fields of $\simeq10^6$\,G in a light cylinder.
\end{abstract}}
\end{center}\vspace*{-0.6cm}

\begin{center}
\parbox{16.5cm}
{\bf\jiuhao neutron star, pulsar, radiation mechanisms, instabilities in plasmas}
\end{center}

\begin{center}
%{\PACS{\rm 23.40.-s, 23.40.Bw}}%・?àào?
%\CITA    %%(11??×￠êí￡oCitation?úèY×??ˉéú3é)
%

\Cit{W. Y. Wang, Pulsar giant pulse: coherent instability near light cylinder, Sci. China-Phys. Mech. Astron. xx, xx (2018), ...}%%(11??×￠êí￡oCitation ?úèYDèê??ˉì?D′)
\end{center}

\textwidth=178truemm \textheight=236truemm%%%%%%D?°?ê?òa?óé?

%%%%%%%%%%%%%%%%%%%%%%%%%%%%%%%%%%%%%%%%%%%%%%%%%%%%%%%%%%%%
\wuhao\vspace*{1.5mm}

\begin{multicols}{2}

%%%%%%%%%%%%%%%%%%%%%%%%%%%%%%%%%%%%%%%%%%%%%%%%%%%%%%%%%%%%
%% Text of article.
%%%%%%%%%%%%%%%%%%%%%%%%%%%%%%%%%%%%%%%%%%%%%%%%%%%%%%%%%%%%
%    Section headings
\renewcommand{\baselinestretch}{1.08} \baselineskip 12.2pt\parindent=10.8pt

\renewcommand{\thefootnote}

%\noindent
%Dear Editors,

%\vspace{2mm}
\noindent
\section{Introduction}
Giant pulse (GP), which is a special form of pulsar radio emission, is a burst-like individual radio pulse obtained from pulsars \cite{Kuz07}. Their  observed flux densities are tens or hundreds of times or even much larger  than those of the average pulse (AP), which differ from the flux densities of regular individual pulses that do not exceed a factor of 10. These individual pulsed emissions carry important and detailed information related to the physics of pulsar radio emission.
The first GP was detected from the Crab pulsar (PSR B0531+21), which is considered to be a remarkable GP emitter \cite{sta68}. The GPs in the Crab pulsar were mainly detected in the following two phases: main pulse (MP) and interpulse (IP) in the frequency ranging from 20\,MHz to 46\, GHz \cite{cor04,crab3,crab1}.
Hankins et al. \cite{han03} observed an extremely narrow nanosecond structure, which is a subset of GP and which corresponds to a high degree of circular polarization. The distribution of the energy fluxes of GP results in a power law distribution with a spectral index of $-2.0$ to $-4.2$ \cite{pop07}, whereas that of a normal pulse is considered to be Gaussian or log-normal distributed \cite{hes74,rit76}. This condition indicates that GPs may exhibit a different emission mechanism from that of the normal pulses. Besides the Crab pulsar, similar characteristic features of GPs can also be observed in a Crab-like pulsar, PSR B0540$-$69, which is a young supernova-remanent  central pulsar \cite{0540}, as well as in several millisecond pulsars (MPs, e.g., PSR B1937+21, \cite{1937b}). A common property of these host pulsars is the high magnetic field of $B_{\rm LC}\simeq10^6$\,G in a light cylinder (LC).
Furthermore, GPs can be discovered in pulsars with $B_{\rm LC}$ values ranging to a few hundred Gauss \cite{1112,1752} . Here, GPs can be classified into the following two types (e.g., \cite{Kuz07}): pulsars with $B_{\rm LC}\simeq10^6$\,G (type I) and those with $B_{\rm LC}\sim10-100$\,G (type II).

Further, various studies have been conducted to understand the generation of GPs. For example, a type of GP, nanoshot, was predicted using a plasma wave turbulence model \cite{wea98}. The radiation around the plasma frequency is linearly polarized along the magnetic field lines, whereas nanoshots from the Crab pulsar can exhibit strong circular polarization \cite{eil16}.
Petrova \cite{pet04} argued that GPs are caused because of induced Compton scattering of the radio radiation off the plasma of the magnetosphere. The frequency spectrum should be continuous after the emission wave is scattered by the plasma of pulsar magnetosphere. However, spectral band structures are observed in the IP of the Crab pulsar \cite{han07}, which are reminiscent of the ``zebra bands'' observed in type IV solar flares \cite{slo72}.
Alternatively, GPs are proposed to have originated from anomalous cyclotron resonance on the last closed magnetic field line near LC \cite{Lyu07}. The first few harmonics approximately match the bands at $\nu\simeq6-10$\,GHz, whereas high harmonics are observed to deviate at bands with higher frequencies. Band spacing shows a linear function of $\Delta\nu\simeq0.06\nu$ from 5\,GHz to 30\,GHz, which can rule out various models \cite{han16}. These band structures are assumed to be generated from the source itself or because of some propagation effects, except for stellar scintillations, which denote that $\Delta\nu\propto\nu^{4-4.4}$ (e.g., \cite{Lam99}).
In this study, we suggest that the Crab pulsar GP is caused by the coherent instability of plasma near the magnetic equator of LC. The radiative waves are amplified at frequencies close to the electron cyclotron harmonics owing to cyclotron instability. Further, the resonance between the cyclotron-resonant-excited wave and background plasma oscillation increases the coherent instability.
In Section 2, we explain the formation of the radiative regions and the emission process. In Section 3, we apply our model to Crab pulsar. Further discussions are presented in Section 4; finally, a summary is provided in Section 5.

\section{Coherent instability at near LC}
\subsection{Emission region at near LC}

\begin{table*}
\begin{center}
\caption{Summary of GPs}
\begin{tabular}{cccccc}
\hline \hline
PSR & P (s) & Frequency (MHz) & $B_{\rm{LC}}\,(\rm{G})$ & $S_{\rm GP}/S_{\rm avg}$ & Reference \\
\hline
B0031$-$07 & 0.9430 & 40,111 & 6.87 & 400 & \cite{0031}\\
J0218$+$4232 & 0.0023 & 610 & $3.14\times10^5$	 & - & \cite{0218}\\
B0301$+$19 & 1.3876  & 111 & 4.67 & 69 & \cite{0301}\\
B0531$+$21(Crab) & 0.0334 & 20-30000 & $9.35\times10^5$ & 50000 & \cite{crab4,crab2}\\
 & & & & & \cite{crab3,crab1}\\
B0540$-$69 & 0.0506 & 1380& $3.53\times10^5$ & $\gtrsim5000$ & \cite{0540,joh04}\\
B0643$+$80 & 1.2144 & 103 & 11.1 &- & \cite{0643} \\
B0656$+$14 & 0.3849  & 111 & 7.49 & 630 & \cite{0656}\\
B0950$+$08 & 0.2531 & $39-112$ & 138 & 490 & \cite{0950a,0950b}\\
B1112$+$50 & 1.6564 & 111 & 4.15 & 80 & \cite{1112}\\
B1133$+$16 & 1.1879 & 111 & 11.7 & $86$ & \cite{1133}\\
B1237$+$25 & 1.3824  & 111 & $4.05$ & 65 & \cite{1237}\\
J1752$+$2359 & 0.4091 & 111 & 69.6 & 320 & \cite{1752}\\	
B1820$-$30A & 0.0054 & 6850 & $2.47\times10^5$ & 1700 & \cite{1820}\\
B1821$-$24 & 0.0031 & 1510 & $7.25\times10^5$ & - & \cite{1821}\\
B1937$+$21 & 0.0016 & 111-5500 &	$9.93\times10^5$ & 600 & \cite{1937a,1937b}\\
B1957$+$20 & 0.0016  & 610 & $3.68\times10^5$ & - & \cite{0218}\\
\hline \hline
\end{tabular}
\label{tab1}

{\bf Notes.} Here PSR is a pulsar name, $P$ is period and $S_{\rm GP}/S_{\rm avg}$ is an excess of the peak flux density of a strongest GP over the peak flux density of an AP.
\end{center}
\end{table*}

\begin{figure}[H]
\begin{center}
\includegraphics[width=0.48\textwidth]{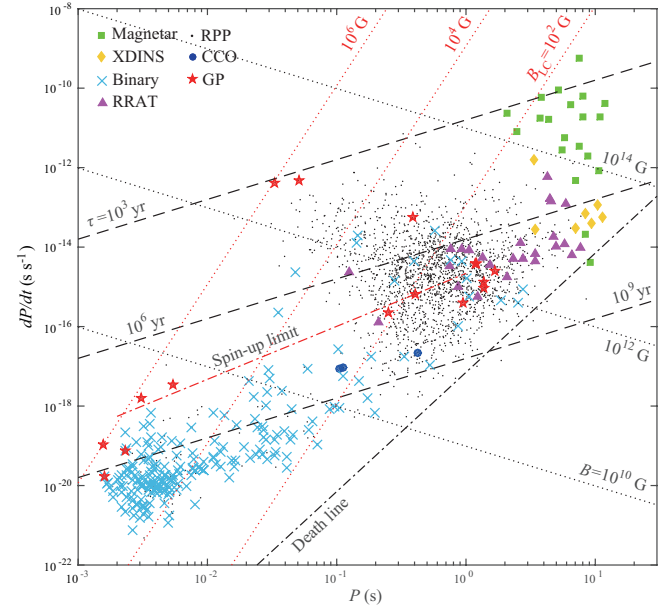}
\caption{\small{(Color online) $P$-$\dot{P}$ diagram of pulsars, including pulsars which have been detected GPs (red stars), rotation-powered pulsar (black points), magnetars (green squares), X-ray-dim isolated neutron stars (yellow diamonds), central compact objects (blue circles), rotating radio transients (magenta triangles), and pulsars in binaries (light blue crosses). The pulsar population data are from ATNF Pulsar Catalogue \cite{atnf}. The death line for a typical $R = 10$\,km neutron star is indicated by the black dashed-dotted line $BP^{-2}=1.7\times10^{11}\,\rm{G\,s^{-2}}$, \cite{Bha92}. The spin-up limit is shown as the red dashed-dotted line $P=1.9(B/10^9\,\rm{G})^{6/7}$\,ms \cite{van87}.}}
\label{fig1}
\end{center}
\end{figure}

\begin{figure}[H]
\begin{center}
\includegraphics[width=0.48\textwidth]{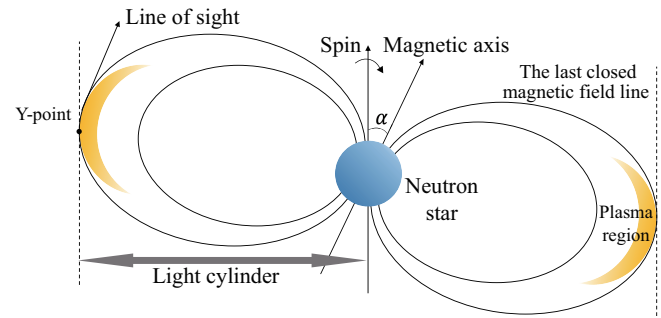}
\caption{\small{Emission region of type I GP. The beamed pair plasma is blocked near LC, co-rotates and form an emission region (yellow zone).}}
\label{fig2}
\end{center}
\end{figure}

We summarize the comparative data for all known GP host pulsars in Table \ref{tab1}.
Based on the distribution of GP sources in Figure \ref{fig1}, we propose that the type I GP emission comes from the plasma at near LC.
The Crab pulsar is a typical type I GP emitter.
Generally, for a rotating magnetized neutron star, a strong electric field can be induced at the magnetosphere \cite{gol69}, producing an electron-positron($e^{-}-e^+$) pair plasma.
Secondary $e^{-}-e^+$ pairs are produced due to charged particles which are accelerated to $\gtrsim10^{11}$\,eV.
And more pairs are produced until a pair avalanche is formed (e.g., \cite{rud75}).
These avalanche-like pair productions lead to the plasma density in open field lines much higher than the Goldreich-Julian(G-J) density.
Such processes hardly happen in closed field line region in which charge density is equal to their local G-J density.
The G-J density is $n_{\rm GJ}=-[{\boldsymbol{\Omega\cdot B}}/(2\pi ce)][1-(\Omega R/c)^2\sin^2\Theta]^{-1}$, where ${\boldsymbol \Omega}$ denotes the rotation vector of the neutron star and $\Theta$ measured from the rotation axis.
It may be significantly enhanced in the current layer and sheet that makes a dense plasma region formed near LC in the equatorial plane.

Magnetic reconnection occurs at the internal edge of the current sheet where it comes to the closed lines of the magnetosphere.
This process loads the open field lines with the plasma and causes field-aligned accelerating fields \cite{aro08}.
Charged particles ($e^--e^+$ pair plasmas) are strongly accelerated in the reconnection region and injected into closed field line regions via the Y-point (e.g., \cite{lyu96,Ist04,zen07}).
The injected particles are beamed and ultra-relativistic that brings much free energy to generate plasma instability.
The maximum Lorentz factor of the electrons acquired from the reconnection was estimated to be $\gamma_{\rm max}\sim10^{11}$ in the polar magnetosphere \cite{Ist04}.
Basically, the maximum electron energy is limited by radiative damping.
With the assumption of curvature-radiation-dominant loss process, the Lorentz factor of fast particles on closed lines is estimated to be $10^7$ by \cite{Lyu07}.

Magnetic field lines close to LC are shaped like a magnetic mirror.
Particles moving along the field lines into the neutron star experience an increasing force that eventually causes them to reverse direction and return to the confinement area.
As a result, particles acquired larger velocity which is perpendicular to the magnetic field through the reconnection, are easier to be trapped and form a banana-like radiation region, shown in Figure \ref{fig2}.
The inclination angle for the Crab pulsar is suggested to be $\alpha=45^\circ$ \cite{du12}.
The radiation region is co-rotating with its host pulsar because of the effect of frozen-in field lines.
Also, radiation of the beam particles makes them cool. 
These cooled particles can be thought of as the background plasma which is fed by the trapped beam particle.
Therefore, particles at near LC in the closed field line region could be regarded as two different parts: nonthermal beam particles and the cold background plasma.

\subsection{Microburst with band structure}
\subsection{Coherent instability with band structure}

It is proposed that the Crab pulsar GP originated from coherent instability of ultra-relativistic pair plasmas near LC.
Beam particles can bring much free energy that produces cyclotron resonance instability.
Electron cyclotron resonance instability, e.g., electron cyclotron maser, was first proposed by \cite{Twi58} who considered that the induced absorption may be negative for electrons with a population inversion in a higher energy state.
It is a significant process that generates coherent radiation from magnetized plasmas (e.g., the Sun, other stars and magnetized planets, \cite{Tre06}).
Here, this process is argued to be generated on a pulsar with a strong magnetic field and an ultra-relativistic plasma beam.
Kazbegi showed that the conditions for the cyclotron instability development are satisfied at near pulsar LC length scales \cite{kaz92}.

The high-frequency electromagnetic waves could be amplified by a resonant interaction between
the wave and energetic electrons at the Doppler shifted electron-cyclotron frequency and its harmonics \cite{Twi58,Bek61}.
This resonant process could occur at highly excited states rather than ground state of the Landau levels.
The phase velocity can exceed the speed of light in the plasma.
Then, the general cyclotron resonance condition for harmonic number $s$ ($s>0$, i.e., normal cyclotron resonance) is
\begin{equation}
\omega-k_{\parallel}v_{\parallel}=s\frac{\omega_{B}}{\gamma_b},
\label{eq1}
\end{equation}
where $\omega$ is the frequency of the excited wave, $\gamma_b$ is the Lorentz factor of electron beam, $k_{\parallel}$ and $v_{\parallel}$ are components of the wave-vector's and particle's velocity along magnetic field, and $\omega_{B}=eB/m_{\rm e}c$ is the cyclotron frequency of electrons, in which $m_{\rm e}$ is the mass of electron and $B$ is the strength of magnetic field.

Growth of the instability occurring for the resonant electrons lie in regions of $\partial f/\partial p_{\perp}>0$ i.e., loss cone, where $p_{\perp}$ is the electron momentum perpendicular to the magnetic field and $f$ is the distribution function of electrons, because the gradient in $f$ provides the free energy which drives the instability.
Wu discussed that the case of resonance ellipses for $k_{\parallel}c<\omega<s\omega_{B}$ in velocity space \cite{wu85}.
With a strong magnetic field, the resonance ellipse is situated on the original point that makes a small fraction of the ellipse inside the loss cone.
Therefore, it is difficult to provide much free energy to induce the instability.
However, for ultra-relativistic particles, one should consider the resonance ellipse in momentum space.
Equation (\ref{eq1}) in momentum space can be written as the formula of a resonance ellipse,
\begin{equation}
(\frac{p_{\perp}}{m_{\rm e}c})^2+(1-\frac{k^2_{\parallel}c^2}{\omega^2})[\frac{p_{\parallel}}{m_{\rm e}c}-\frac{s\omega_{B}k_{\parallel}c}{\omega^2-k^2_{\parallel}c^2}]^2=\frac{s^2\omega^2_{B}}{\omega^2-k^2_{\parallel}c^2}-1.
\label{eq2}
\end{equation}
The ellipse is situated on one side of the origin that may make a larger portion of the ellipse inside the loss cone.
It makes much more particles can provide free energy that triggers the instability.

When radiation propagates across the magnetic field, there are two eigenmodes: the extraordinary mode (X-mode, the electric field is perpendicular to the $k$-$B$ plane) and the ordinary mode (O-mode, in the plane).
The dispersion relation in the magnetized plasma can be written as \cite{wu02},
\begin{equation}
\begin{split}
N^{2}_q=\frac{k^2c^2}{\omega^2}=1-{\Omega_{p,b}^2\over \omega\left(\omega+\tau_q\Omega_{B,b}\right)},\\
\tau_q=-s_q+q\sqrt{s_q^2+\cos^2\theta},\\
s_q=\frac{\omega\Omega_{B,b}\sin^2\theta}{2\left(\omega^2-\Omega_{p,b}^2\right)},
\label{eq3}
\end{split}
\end{equation}
where $N_q$ are the refractive index, $q=\pm$ denote the O and X modes, $\Omega_{B,b}=\omega_{B}/\gamma_b$ is the observed cyclotron frequency of relativistic electron beam, $\theta$ is the angle between the ambient magnetic field vector and the wave vector, and $\Omega_{p,b}=\sqrt{(8\pi n_be^2)/(\gamma_b m_{\rm e})}$ is the total plasma frequency of the pair plasma beam, in which $n_b$ is the number density of the electron beam.
The polarization is given by \cite{gin70},
\begin{equation}
\begin{split}
K_q=\frac{e_y}{e_x}=-i\frac{\cos\theta}{\tau_q+2s_q},\\
T_q=\frac{e_z}{e_x}=\frac{i\omega\Omega_{B,b}\Omega^2_{p,b}\sin\theta-\Omega^2_{B,b}\Omega^2_{p,b}K_q\sin\theta\cos\theta}{\omega^4-\Omega^2_{B,b}\omega^2-\Omega^2_{p,b}\omega^2+\Omega^2_{B,b}\Omega^2_{p,b}\cos^2\theta},
\label{polar}
\end{split}
\end{equation}
where $z$ axis is parallel to the wave-vector.
Both O and X modes exhibit elliptic polarization: X mode is right-hand polarized, whereas O mode is left-hand polarized.
For these ultra-relativistic plasmas, their emission propagates approximately along the direction of magnetic field because of the beaming effect.
In the limit of $\theta\sim1/(2\gamma_b)\ll1$, $\Omega_{B,b}\ll\omega$ and $\Omega_{p,b}\ll\omega$, the cut-off frequency for the two modes are \cite{mel17}
\begin{equation}
\omega_{\rm L(R)}=\frac{-q\Omega_{B,b}\cos\theta+\sqrt{4\Omega^2_{p,b}+\Omega^2_{B,b}\cos^2\theta}}{2},
\label{eq10}
\end{equation}
where $\omega_{\rm L}$ for the left-hand O mode and $\omega_{\rm R}$ for the left-hand X mode.
From equation (\ref{eq1}) and (\ref{eq3}), one can obtain,
\begin{equation}
\omega=\frac{\Omega^2_{p,b}}{2s\Omega_{B,b}}\mp\Omega_{B,b}.
\label{eq8}
\end{equation}
Thus, the spectrum exhibits zebra-pattern-like band structures during the cyclotron instability occurs.

A successful model for the solar zebra bands is the double plasma resonance, which proposed that enhanced excitation of plasma waves occurs at resonance levels where the upper hybrid frequency coincides with an integer multiple of the cyclotron frequency\cite{zhe75}.
However, unlike the solar double plasma resonance, the wave resonate with the upper hybrid wave of the $e^--e^+$ pair tail is the cyclotron-resonant-excited wave.
In a magnetized plasma, there is a restoring force due to the Lorentz and the electrostatic Coulomb force.
The oscillation frequency is the upper hybrid frequency $\omega_{\rm uh}$ that can be regarded as the background frequency.
A resonance between the cyclotron-resonant-excited wave and the cold background plasma oscillation would occur when,
\begin{equation}
\omega=\omega_{\rm uh}=\sqrt{\Omega^2_{p,c}+\Omega^2_{B,c}},
\label{eq4}
\end{equation}
where $\Omega_{B,c}=\omega_{B}/\gamma_c$ is the observed cyclotron frequency of the background plasma, $\Omega_{p,c}=\sqrt{(8\pi n_ce^2)/(\gamma_c m_{\rm e})}$ is the total plasma frequency of background plasma, $n_c$ and $\gamma_c$ are number density and the Lorentz factor of the background plasma.

A set of harmonics would be excited when the spatial structure of $n_c$ and $B$ allows this resonance to be satisfied at more than one location.
In the usual polar coordinates $(R, \Theta)$, small changes in $\Omega_{p,c}$ and $\omega_{B}$ at location of $R+\Delta R$ can be written as \cite{wing86},
\begin{equation}
\frac{\Omega_{p,c}(R+\Delta R)}{\omega_B(R+\Delta R)}\simeq\frac{\Omega_{p,c}(R)}{\omega_B(R)}[1-(\frac{1}{L_B}-\frac{1}{2L_n})\Delta R],
\label{eq6}
\end{equation}
where $L_n=n_c(dn_c/dR)^{-1}$ and $L_B=B(dB/dR)^{-1}$ are the scale lengths of the inhomogeneities in $n_c$ and $B$, respectively. 
Here, we assume that $n_b(R)$ has same functional form as $n_c(R)$, i.e., $n_b(R)=\kappa n_c(R)$, where $\kappa$ is a constant as well as $\gamma_c$.
For $\Omega_{B,c}\ll\Omega_{p,c}$, the location change is given by,
\begin{equation}
|\Delta R|=\frac{1}{s}\left|\frac{1}{L_B}-\frac{1}{2L_n}\right|^{-1}.
\label{eq0}
\end{equation}
Then, for high harmonics, the stripe frequency separation is given by
\begin{equation}
%\begin{split}
\Delta \omega_{s,s+1}=\omega_s-\omega_{s+1}\simeq\frac{2\omega_s\omega_B}{\Omega_{p,c}\gamma_c\kappa(\frac{2L_n}{L_B}-1)}.
\label{eq7}
%\end{split}
\end{equation}
Therefore, band structures are modulated by this wave-wave interaction.
The stripe frequency separation of the modulated bands is proportional to the band-center frequency.
 
\section{GP in the Crab pulsar}

\begin{figure}[H]
\begin{center}
\includegraphics[width=0.48\textwidth]{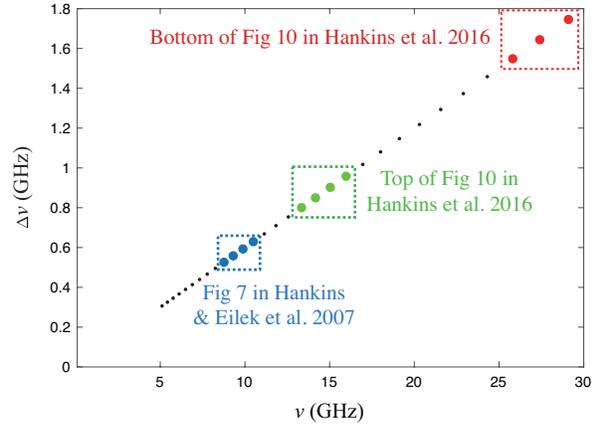}
\caption{\small{Band structure of the Crab pulsar. With $\omega_B=8.4\times10^3$\,GHz, and assumption of $\kappa=1$ and $L_n=1.5L_B$, dots are modeled band-center frequencies at different harmonics. Red and green dots are bands from Figure 10 in \cite{han16}. Blue dots are bands from Figure 7 in \cite{han07}. The observed band structures can be well fitted by equation (\ref{eq7}).}}
\label{fig4}
\end{center}
\end{figure}

The Crab pulsar is the most active and a bright GP emitter which has been detected.
Most MPs and low-frequency IPs are made up of one to several microbursts.
These microbursts can occur anywhere in the probability envelope defined by the mean-profile component with variable strength \cite{eil16}.
Also, they can often be resolved into short-lived, overlapping nanoshots \cite{han03,han07}.
In contrast, nanoshots have not been found in high-frequency IPs \cite{han16}.
Both microbursts and nanoshots are GPs.
Spectral bands are found in microbursts of IPs at multi-frequencies.
Band spacing measurements of IPs in the range of $5-30$\,GHz are well fitted by a linear function of $\Delta \nu=0.06\nu$.
It can be well fitted by equation (\ref{eq7}), shown as Figure \ref{fig4}.

For the Crab pulsar, one can calculate the cyclotron frequency $\omega_B\simeq8.4\times10^3$\,GHz at near LC.
From equation (\ref{eq4}), it is required that $n_b>8.4\times10^{13}\,\rm{cm^{-3}}$ in order to have frequency $\gtrsim5$\,GHz.
A typical timescale of IP spectral band structures is several micro-seconds which indicates an emitting source scale of $\sim3\times10^4$\,cm.
The coherence scale of the cyclotron instability is supposed to be the gyroradius of electrons.
The Lorentz factor of beam particles can be estimated to $\gamma_b\sim10^6$.
Hence, one can obtain the lower limit of the cut-off frequency of $\sim0.1$\,GHz.
Band structures of the microburst spectrum are excepted to be detected at lower frequencies.

A combined term $n_c\gamma_c m_{\rm e}c^2\lesssim B^2_{\rm LC}/(8\pi)$ is given by the condition for the background plasma blocked in the magnetic field lines.
The number density of the background plasma at the Y-point can be estimated to $n_c\sim10^{13-15}\,\rm{cm^{-3}}$ with $\gamma_c\simeq10^{2-4}$ in order to make the resonance between the cyclotron-resonant-excited wave and the background oscillation occur.
Such high dense plasma for G-J density is supposed to be within $10^{3-5}$\,cm at near the magnetic equator of LC.
Also, the number density and the Lorentz factor of the $e^{-}-e^{+}$ pair tail produced at open field lines are $10^{4-5}$ times than G-J density of the star surface i.e., $n_t\simeq10^{17-18}\,\rm{cm^{-3}}$ with $\gamma_t\simeq10^{3-4}$ \cite{mac89,hib01}.
A possible reason for the relativistic dense plasma in closed lines near LC is that torsional Alfv$\acute{\rm e}$n waves or pumping magnetic helicity lead to the pair tail particles diffusing into the closed field line region, e.g., the density of PSR J0737$-$3039 on the closed field line exceeds the G-J density by a factor of $10^{4-5}$ \cite{Lyu05}.
Here,  type I GP is supposed to be the performance of high density plasma activities at near LC.

A value of $2L_n/L_B\lesssim3.6$ can be obtained from equation (\ref{eq7}).
In the frame of a dipole magnetic field, $dB/dR$ at near LC can be calculated to $\sim 3B_{LC}/R_{\rm LC}\sim0.1\,\rm{G\,cm^{-1}}$.
The scale lengths of the emission region is approximately equal to $L_n$
The inhomogeneity scale of $n_c$ is $|L_n|<1.8\times10^7$\,cm that indicates an estimation of $|dn_c/dR|>5.6\times10^5\,\rm{cm^{-4}}$.
The absorption could be neglected due to the small length. 
$n_c$ and $B$ are increasing from the position of radiation source along the line of sight.
Thus, high harmonics with low frequency come from the location closer to LC.

The growth of the instability is supported by the free energy that comes from the non-equilibrium anisotropic distribution of fast particles.
One way to provide free energy is that the magnetic reconnection and fast beam particle injection at the Y-point.
The temporal growth rate of the cyclotron instability can be estimated as \cite{lom83},
\begin{equation}
\Gamma_c\simeq \omega_{B}\gamma_0\frac{n_b}{n_c}(\frac{\Omega_{p,b}}{\omega_B})^4\gamma_p^{-2}\gtrsim10^2\,\rm{s^{-1}},
\label{eq5}
\end{equation}
where $\gamma_0\simeq10^{4-5}$ is the mean Lorentz factor of the plasma system on average and $\gamma_p\simeq10^{-5}c/\Omega\simeq10^3$, in which $\Omega$ is the angular velocity of spin. 
The growth rates of different bands are nearly the same.
The observed intensity also depends on saturation mechanism.
In the case of $\Omega_{B,b}\ll\Omega_{p,b}$, the growth rate of low harmonics is faster than that of high (e.g., \cite{mel84}).
The maximum instability growth of loss-cone is proportional to $s^{-2}$ (e.g., \cite{rose65}).
It is consistent with the result that the intensity of high frequency is larger.
The cyclotron instability may grow at a timescale of several milliseconds which meets time of the beam particle escape from the LC.
Basically, the background resonance also helps with the instability increasing.
The final growth rate needs to consider contributions from cyclotron instability and the background-oscillation resonance.

Most MPs consist of several microbursts which can be resolved into nanoshots.
Nanoshot is another genre of GP.
These nanoshots can occur anywhere inside the MP phases with randomly circular polarization \cite{han03}.
The normal modes become circularly polarized for propagation close to the magnetic field: left-hand for $e^-$ moving faster than $e^+$ whereas right-hand for $e^+$ moving faster than $e^-$ \cite{all82}.
There is still no evidence that spectral bands have been observed in nanoshots.
Maybe the frequency resolution is not high enough within several nanoseconds.
In addition, particles tend to drift along the direction perpendicular to the plane containing the curved field lines.
Drift waves can be generated from the drift particles whose velocity is $v_{d}=v_{\parallel}^2|\nabla B|/(B\Omega_{B,b})\approx10^{-4}c$ at near LC.
The velocity of the drift wave is much smaller than that of particle \cite{mac18}.
In the magnetosphere, three and four-wave interaction and nonlinear interaction with plasma particles may exist \cite{gog05}.
The drift wave frequency becomes higher due to the waves merging.
The three-wave probability $W\propto\omega^{-2}$ \cite{luo06} that indicates the induced three-wave interaction is difficult to occur at high frequency bands.
In this scenario, emitting spots form within some small structures where the drift wave accumulates energy through nonlinear wave interaction \cite{mac17}.
The timescale of this process is $\sim1/\Omega_{B,c}\approx1$\,ns which matches the observations.
One can estimate the growth rate of this process \cite{luo94}
\begin{equation}
\begin{split}
\Gamma_n\approx2\Omega_{p,b}(\frac{\gamma_c}{\gamma_b^3})^{1/2}(\frac{\Omega_{p,c}}{\omega})\\
\sim15(\frac{\gamma_b}{10^4})^{-2}(\frac{n_b}{10^{14}{\rm cm^{-3}}})^{-1/2}(\frac{n_c}{10^{13}{\rm cm^{-3}}})^{-1/2}\,s^{-1}.
\end{split}
\end{equation}
In addition, the optical depth is estimated as \cite{luo06}
\begin{equation}
\tau\approx1.2\times10^{-3}(\frac{S_\nu}{10^3\,{\rm Jy}})(\frac{D_{\rm GP}}{1\,{\rm kpc}})^2(\frac{\gamma_c}{10^4})^{-2}(\frac{r}{10^8\,{\rm cm}})(\frac{n}{n_{\rm GJ}}),
\end{equation}
where $S_\nu$ is the GP flux and $D_{\rm GP}$ is the distance from us.
It is optical thin so that GP photons can escape from the LC.

\section{Discussion}
\subsection{Radiation at near LC}
Normal radio pulse emissions for many pulsars are thought to be originated from lower altitudes or polar caps.
However, the Crab pulsar may have different emission geometry that both normal radio pulse and GP may be derived from the region near LC.
Peaks of optical, X-ray and $\gamma$-ray pulse components have precisely the same pulse phases as GPs in the Crab pulsar \cite{han07}.
Moreover, in millisecond pulsars (e.g., \cite{cus03}), GPs are closely aligned in phase with the pulse component observed at X-ray bands.
For instance, in PSR J0218+4232, GPs are coincident with the phases of the X-ray peaks whereas they deviate from peaks of mean radio pulse profile \cite{kni06}.
Pulse phases coincidence of radio GP and high energy emission is interpreted by their same emission area.

In the Crab pulsar, the upper limit of luminosity for high energy emission is $\sim B^2_{\rm LC}R^2_{\rm LC}c\simeq3\times10^{38}\,\rm{erg\,s^{-1}}$ due to the magnetic reconnection.
Spectrum of $10\,\rm{MeV}-2.5\,\rm{GeV}$ are interpreted by synchrotron emission near LC \cite{chk11}.
Inverse Compton scattering from collisions between $e^-/e^+$ and soft photons, and curvature radiation from $e^-/e^+$, are possible ways in generating $\gtrsim1\,\rm{GeV}$ radiation.
Pair annihilation in the GP emission region can create $\lesssim0.5\,\rm{MeV}\,\gamma$-rays.
The spectrum of these $\gamma$-rays may be power-law like because cross-section for $e^--e^+$ is proportional to $\gamma^{-2}$.
Moreover, $e^--e^+$ pairs produced in the current sheet near LC emit in near infrared and optical band which have same pulse phases as radio GP \cite{lyu96}.

The emission regions are patchy and dynamic that leads to the radio power which fluctuates on a wide range of timescales \cite{eil16}.
Non-uniformity of the emission region is a necessary condition for a set of harmonics excited by the background plasma oscillation resonance, while inhomogeneities of the emission region may broaden the stripes.
The number density gradient is suggested to be very large, that makes emissions generation and resonance with the background oscillations occur at extremely closed locations.
On the other hand, if the radiation region is more internal to the magnetosphere, it may be significantly absorbed by particles in the magnetosphere.
Photons are much easier to escape near LC.

The loss of electrons with small pitch angles due to the precipitation into a high density ``corona'', can give rise to a loss-cone distribution in the momentum space.
The plasma region meets the condition of the loss-cone momentum distribution.
There are many kinds of loss-cone distribution functions for relativistic electron plasmas, e.g., \cite{pri84,tsang84}.
Growth of the normal cyclotron resonance can be generated when resonant electrons lie in a loss-cone.
This process is different from cyclotron-Cherenkov resonance which is a wave-wave interaction \cite{Lyu99,Lyu07}.
Similar with pulsar GP, zebra patterns are also observed in solar radio burst.
Microwave zebra patterns associated with solar flares can be classified into three types in which the growing-distance pattern can be explained by the double plasma resonance \cite{tan14}.
However, if the GP band structures are caused by double plasma resonance, a much lower magnetic field is needed.

Generally, the instability is caused by beam particles which are powered by the reconnection at near LC.
A strong magnetic field can provide enough free energy that may be the reason why type I GP tends to occur at pulsar with $B_{\rm LC}\simeq10^6$\,G. 
Therefore, more GP events are expected to be found in high $B_{\rm LC}$ pulsars.
The band structure may be a common property among all type I GP source.
Future advanced facilities may provide unique opportunities to search more samples and understand more informations about GP.
We are looking forward to detect these structures in other type I GP source and that in lower frequencies of the Crab pulsar.

\subsection{GP sources}
It is worthy to note that GPs are multi-originated.
In the pulsar $P$-$\dot{P}$ diagram, shown as Fig \ref{fig1}, the sources with GP are divided into two parts.
The separation of the GP sources in the $P$-$\dot{P}$ diagram suggests that type I and type II GPs are come from different regions.
For instance, Kuzmin found that the frequency dependence of the separation of GP emission regions for PSR B0031$-$07 is similar to that of the width of the AP \cite{0031}.
This suggests that the type II GPs are emitted from a hollow cone over the polar cap instead of LC.
Type II GPs are also found in PSRs B1133+16 and B1112+50.
The fact of their pulse phase ranges narrower than that covered by the mean pulse profile is similar with cases of the Crab pulsar and PSR 1937+21 \cite{kar11}.
However, there are some differences between type I and type II GP.
Typical timescale of type II GP is several milliseconds while that of type I is micro-second or even nano-second.
The maximum flux density of type II GP rarely exceeds that of AP by a factor of a thousand.
Phase of type II GP is stable inside the integrated profile whereas type I GP can occur at phase outside mean pulse profile \cite{1937b}.
Thus, type II GP is also considered as a so-called ``not-so-GP'' \cite{man09}.
Additionally, type II GP sources are located with rotating radio transients (RRATs).
The pulse intensity distributions of RRATs are log-normal with power-law tails and the spectra is power-law like \cite{mil11}.
These characteristic features meet normal pulses.
Type II GPs are likely to be some bright normal pulses.

For type I GP sources, the Crab pulsar and PSB B0540$-$69 are located away from other sources some of which are milli-second pulsars.
A common feature of these sources is the high magnetic field at near LC, shown in Figure \ref{fig1}.
MP and IP of 1.4 GHz appear at the same phase as two peaks in the high energy light curve that is similar to the case in millisecond pulsars \cite{abd10,joh14}.
This may imply that the Crab pulsar and millisecond pulsar have the same emission region.
Both the Crab pulsar and PSR B0540$-$69 are central pulsar of supernova remnant \cite{man93}.
They are surrounded by strong pulsar wind nebulas which are dynamic dense magnetized clouds of electrons.
The backward pulsar wind nebula offers possibilities for the formation of the high density plasma at near LC.
In addition, for millisecond GP sources, some of these are identified as binaries.
Accreting material from a companion star may form the dense plasma region near LC.
In the these cases, maybe binary pulsar or remnant-central pulsar with high magnetic field at LC is easier to detect GP.
Maybe we could observe some sporadic bright individual pulses which are derived from the ``missing'' pulsar in SN 1987A \cite{man07}. 

\subsection{GP and Fast Radio Burst}
Similar with GP, fast radio burst (FRB) is also a phenomenon of radio flash with high flux density.
Many models suggest that FRBs are originated from magnetars (e.g., \cite{pen15,met17}) while there is no evidence that GPs are detected in magnetars.
Recently, rotation measurement shows that the environment of the repeater FRB 121102 is similar with a galactic center magnetar PSR J1745$-$2900 \cite{mic18}.
The linear polarization of the repeater is about 100\% after correcting for Faraday rotation and accounting for about 2\% depolarization from the finite channel widths \cite{mic18}.
The repeater source itself is strong linearly polarized.
This scenario is consistent with high-frequency IP while deviates from MP of the Crab pulsar.
Also, over half of the peaks, GPs are known to be approximately 100\% circularly polarized of PSR B1937+21.
The polarizations of the repeater FRB do not show any significant change \cite{gajjar}, while those of GP are variable.
The differences in polarization between the repeater and GP may suggest that they have different origins.

The narrow frequency width of GP can be estimated as $\Delta\nu\simeq0.06\nu\sim1$\,GHz that is assumed to be equal to that of FRB.
The maximum flux of the Crab GP is $S_{\rm GP}\sim1$\,MJy at a distance of $d_{\rm Gp}\simeq1$\,kpc.
GPs are repeated events while only FRB 121102 has been found to be repeatable.
The repeater has a few hundred milli-Jansky flux with typical distance $d_{\rm FRB}\simeq 1\, \mathrm{Gpc}$ (\cite{cha17}).
With these parameters, one can obtain the ratio of their luminosities,
\begin{equation}
\frac{L_{\rm FRB}}{L_{\rm GP}}=\frac{S_{\rm FRB}d^2_{\rm FRB}}{S_{\rm GP}d^2_{\rm GP}}\sim10^5.
\label{eq9}
\end{equation}
The luminosity of FRB is $10^{41-44}\,\rm{erg\,s^{-1}}$ \cite{luo18} much higher than that of GP.
Alternatively, this phenomenon may be caused by the selection effect.
The power-law index of cumulative distribution for a repeating FRB sequence is $-1.16$ \cite{wang18} steeper than that of GP.
In the this case, FRB can be regarded as a flatter high energy tail of GP possibly.
A burst width can be broaden due to multipath propagation through the interstellar medium.
This phenomenon may be very significant for distant radio sources.
For the repeater FRB in $4-8$\,GHz, the broadening time $\tau_{d}$ is calculated to be $0.01-0.1$\,ms (see \cite{bha04}, for the test of 5\,GHz and DM$=565\,{\rm pc\,cm^{-3}}$ ).
Hence, the intrinsic width of the repeater FRB is about few milliseconds which is different from the case of type I GP.
Type II GP has a type width of $\sim1\,$ms whereas the maximum flux is much smaller than that of type I GP.

GP and the repeater FRB exhibit different properties among polarization, luminosity and timescale. 
Recently, band structures are detected in the repeater with band spacing of $1-100$\,MHz at $4-8$\,GHz, while no evidence of $\Delta\nu\propto\nu$ are found \cite{gajjar}.
These band stripes in the repeater are speculated to be generated from interstellar scintillations.
In general, it is suggested that the repeater FRB is not a pulsar GP.

\section{Summary}
Based on the different properties of GP host pulsars, we propose GPs can be identified as two kinds: magnetic field at light cylinder with a factor of $10^6$ G (type I) and that the case of few ones to a several hundreds (type II). 
These two type GPs exhibit differences among timescale, energy distribution and occurrence phase.

We note that the Crab pulsar GP could be originated from the coherent instability of plasmas at near LC.
The magnetic reconnection could occur at here and accelerate pair plasmas injecting into the closed field line region.
Since the magnetic mirror like shape of field lines closed to LC, the injected particles may be trapped forming a banana-like emission region.
These beam particles bring significant free energy that makes cyclotron resonant instability grows and amplifies radiative waves at frequencies close to the electron cyclotron harmonics.
The spectrum of the cyclotron-resonant-excited wave shows zebra-pattern-like spectral band structures.
These structures can be modulated by the resonance between the cyclotron-resonant-excited wave and the background plasma oscillation.
The linear band spacing in IP of the Crab pulsar can be well fitted by the model of coherent instability at near LC. 
The modeled density of plasmas at LC is $10^{13-15}\,\rm{cm^{-3}}$ with an estimated gradient of $>5.5\times10^5\,\rm{cm^{-4}}$.
Hence, GP is the performance of dense plasmas activities at near LC.
This process may be associated with some high energy emissions.
Similar band structures are expected to be detected in more type I GPs with multi-frequencies.

\noindent
{\small \em
We are grateful to Daming Qu and Dr. BaoLin Tan at National Astronomical Observatories, Chinese Academy of Sciences (CAS), Dr. Duncan R. Lorimer at West Virginia University, Morgantown, Dr. Guoqing Zhao at Luoyang Normal University, and all of the members in the pulsar group at Peking University for comments and discussions about the model.
We are also grateful to Tim Hankins at New Mexico Tech and National Radio Astronomy Observatory for providing data and comments.
Weiyang Wang and Xuelei Chen acknowledge the support of MoST 2016YFE0100300, Natural Science Foundation of China (NSFC) 11473044, 11633004, 11653003, CAS QYZDJ-SSW-SLH017.
Jiguang Lu is supported by the NSFC 11225314 and the Open Project Program of the Key Laboratory of Radio Astronomy, CAS.
Rui Luo is supported by NSFC U15311243 National Basic Research Program of China, 973 Program, 2015CB857101, XDB23010200, 11690024, 11373011, and funding from the Max-Planck Partner Group.
Renxin Xu acknowledges the support of National Key R\&D Program of China (No. 2017YFA0402602), NSFC 11673002 and U1531243, and the Strategic Priority Research Program of CAS (No. XDB23010200)
Baolin Tan acknowledges the support of NSFC 11573039, 11661161015 and 11790301.
The FAST FELLOWSHIP is supported by Special Funding for Advanced Users, budgeted and administrated by Center for Astronomical Mega-Science, Chinese Academy of Sciences (CAMS).}

{}

\end{multicols}

\end{document}